\documentstyle[12pt,epsfig,graphicx,amsmath]{elsart}

\begin{document}

{\tt DESY 00-160}\\
{\tt TPR-00-20  }\\
{\tt November 2000}

\begin{center}
  
\title{\bf Electron Scattering with Polarized Targets \\
        at TESLA}

\collab{\em The TESLA-N Study Group}
\footnote{Contact:Wolf-Dieter.Nowak@desy.de}

\author[torino]{M.~Anselmino},
\author[desyz]{E.C.~Aschenauer}, 
\author[peter]{S.~Belostotski},
\author[desyh]{W.~Bialowons},
\author[desyz]{J.~Bl\"umlein},
\author[regen]{V.~Braun},
\author[desyh]{R.~Brinkmann},
\author[bay]{M.~D\"uren}, 
\author[desyz]{F.~Ellinghaus},
\author[bochum]{K.~Goeke},
\author[bochum]{St.~Goertz},
\author[erlangen]{A.~Gute},
\author[bochum]{J.~Harmsen},
\author[mainz]{D.~v.Harrach},
\author[wupp]{R.~Jakob},
\author[mainz]{E.M.~Kabuss},
\author[desyz]{R.~Kaiser},
\author[desyz]{V.~Korotkov},
\author[wupp]{P.~Kroll},
\author[london]{E.~Leader},
\author[regen]{B.~Lehmann-Dronke},
\author[warsaw]{L.~Mankiewicz},
\author[bochum]{A.~Meier},
\author[bochum]{W.~Meyer},
\author[desyh]{N.~Meyners},
\author[regen]{D.~M\"uller},
\author[ams]{P.J.~Mulders}, 
\author[desyz]{W.-D.~Nowak},
\author[regen]{L.~Niedermeier},
\author[desyh]{K.~Oganessyan},
\author[bochum]{P.V.~Pobilitsa},
\author[bochum]{M.V.~Polyakov},
\author[bochum]{G.~Reicherz},
\author[erlangen]{K.~Rith},
\author[gent]{D.~Ryckbosch},
\author[regen]{A.~Sch\"afer},
\author[desyh]{K.~Sinram},
\author[nikhef]{G.~v.d.Steenhoven}, 
\author[erlangen]{E.~Steffens},
\author[nikhef]{J.~Steijger},
\author[bochum]{C.~Weiss}

\address[ams]{Free University Amsterdam} \vspace*{-0.3cm}
\address[bay]{University of Bayreuth} \vspace*{-0.3cm}
\address[bochum]{Ruhr-University Bochum}\vspace*{-0.3cm} 
\address[desyh]{DESY Hamburg}\vspace*{-0.3cm} 
\address[desyz]{DESY Zeuthen}\vspace*{-0.3cm} 
\address[erlangen]{University of Erlangen-N\"urnberg}\vspace*{-0.3cm} 
\address[gent]{University of Gent}\vspace*{-0.3cm}
\address[london]{Imperial College, London}\vspace*{-0.3cm} 
\address[mainz]{University of Mainz}\vspace*{-0.3cm} 
\address[nikhef]{NIKHEF Amsterdam}\vspace*{-0.3cm} 
\address[peter]{University of St.Petersburg}\vspace*{-0.3cm} 
\address[regen]{University of Regensburg}\vspace*{-0.3cm} 
\address[torino]{INFN Torino}\vspace*{-0.3cm} 
\address[warsaw]{Nicolaus Kopernicus Astronomical Center Warsaw}
\vspace*{-0.3cm} 
\address[wupp]{University of Wuppertal}\vspace*{-0.3cm}

\end{center}

\newpage

{\bf Abstract} 

{\footnotesize \sf
Measurements of polarized electron-nucleon scattering can be realized 
at the TESLA linear collider facility with projected luminosities that 
are about two orders of magnitude higher than those expected 
of other experiments at comparable energies.
Longitudinally polarized electrons, accelerated as a small fraction of the 
total current in the e$^+$ arm of TESLA, can be directed onto a solid state 
target that may be either longitudinally or transversely polarized. 
A large variety of polarized parton distribution and fragmentation 
functions can be determined with 
unprecedented accuracy, many of them for the first time. A main goal of the 
experiment is the precise measurement of the $x$- and Q$^2$-dependence of the 
experimentally totally unknown quark transversity distributions that will 
complete the information on the nucleon's quark spin structure 
as relevant for high energy processes. Comparing their Q$^2$-evolution to 
that of the corresponding helicity distributions constitutes an
important precision test of the predictive power of QCD in 
the spin sector. Measuring transversity distributions and tensor charges 
allows access to the hitherto unmeasured chirally odd operators in QCD which
are of great importance to understand the role of chiral symmetry.
The possibilities of using unpolarized targets and of experiments with a 
real photon beam turn TESLA-N into a versatile next-generation facility at 
the intersection of particle and nuclear physics.}

\tableofcontents

\newpage
%

%=====================
\section{Introduction}         
%=====================
%
\enlargethispage{\baselineskip}
Today there is widespread confidence that Quantum Chromodynamics (QCD) 
is the correct theory of strong interactions. On the level of unpolarized
parton distribution functions the theory has been tested with 
considerable precision by many experiments. However, after 10 years
of intense theoretical and experimental activities in 
studying the {\it polarized} nucleon, the angular momentum composition of 
the nucleon remains a territory with blank spots. 
High precision data in a large kinematic domain are required to fully
explore the spin structure of QCD.

Measurements of polarized deep-inelastic scattering (DIS) were up to now 
mostly performed with longitudinally polarized nucleons. Hence, the majority
of experimental information on the angular momentum composition of the nucleon 
is restricted to its longitudinal spin structure. 
This is characterized through the helicity distributions $\Delta q(x,Q^2)$
(also known as longitudinal quark spin distributions), 
where $q$ denotes the quark flavor, the `Bjorken-variable' $x$
is the fraction of the nucleon momentum carried by the interacting 
parton and  $Q^2$ is the virtuality of the exchanged photon. 
However, of equal importance for a complete understanding of the 
spin structure of the nucleon as seen in high-energy processes, are 
the hitherto unmeasured transversity distributions 
$\delta q(x,Q^2)$, which can only be measured with transversely polarized 
nucleons. 

While a weighted sum of the helicity distributions 
$\Delta q(x,Q^2)$ is directly accessible in inclusive deep-inelastic scattering
(DIS) as the longitudinal spin structure function $g_1(x,Q^2)$, 
the transversity distributions 
$\delta q(x,Q^2)$ do not appear in an inclusive structure function. They can, 
however, be measured in semi-inclusive DIS  
processes which implies a substantially higher experimental effort.
In comparison, the perspectives of RHIC for a direct measurement of 
transversity are not good~\cite{schaefer1}. 

First results on transversity distributions can be expected 
from HERMES \cite{hermes-LRP} and COMPASS \cite{compass-prop} within 3-5 
years from now, while a complete high precision mapping of their $x$- and 
$Q^2$-dependence requires high statistics measurements that are beyond the 
scope of presently or soon running experiments.  

An important reason for the interest in the transversity distributions
and their first moments, the tensor charges, is the fact that these 
quantities are related
to matrix elements of chirally odd operators in QCD \cite{Jaffe:1997yz}. 
All known low-energy probes of hadrons such as electromagnetic or 
weak currents are chirally even, so that low-energy experiments 
cannot provide any information about chirally odd matrix 
elements\footnote{An exception is the so-called sigma term, whose 
effect on hadrons is, however, proportional to the small current quark 
masses.}. Inclusive DIS at large $Q^2$ (both unpolarized and polarized) 
measures only chirally even operators, hence a whole
class of operators so far remained unmeasured because of the lack of
suitable `natural' probes coupling to them. 
Hadrons are expected to react very 
differently to chirally odd probes as compared to chirally even ones; 
e.g. the coupling of the flavor non-singlet tensor charge to pions 
is completely different from that of the axial 
charge \cite{Pobylitsa00}. A measurement of the 
transversity distributions and tensor charges would for the first time 
provide an opportunity to access the `missing' chirally odd operators.
In this way it would greatly improve the understanding of the role of 
chiral symmetry in shaping the structure of the QCD ground state and of
the low-mass hadrons.

The successful understanding and use of {\it un}polarized distribution 
and fragmentation functions in various processes have given confidence 
that QCD can be used not only for the extension to polarized functions.
Moreover, it is also applicable for contributions of higher orders in
the coupling constant $\alpha_s$ or beyond leading order in an
expansion in the inverse hard scale ($Q$ for deep-inelastic
leptoproduction), which is referred to as higher twist.
Progress in these directions requires to develop new calculational techniques 
as well as novel methods to solve the evolution equations associated.
These more involved aspects of QCD are just those that are
perceived by many theorists to be the most interesting ones. 
It is widely accepted that QCD is rich enough as a theory to be 
able to generate the entire hadron and nuclear physics phenomenology.
One crucial aspect, however, for most of the
relevant physics, namely a complete and systematic control 
of all relevant higher-order and higher-twist contributions     
is still not in reach. Straightforward QCD perturbation theory
often only converges for some limited kinematic configurations.
This is in striking contrast to the fact that much of the 
available experimental data is easily interpreted by a combination 
of leading order perturbative calculations and some `intuitive'
power--correction terms. It suggests that
for many signals a QCD-description could be pushed down to photon 
virtualities as low as $Q^2=1$ GeV$^2$. 
High accuracy measurements in this
domain will provide definitive tests for higher-order and higher-twist
QCD calculations.

To the extent that the focus of hadron physics turns to higher energies and 
more exclusive reactions, a corresponding move is about to begin from the 
traditional, somewhat `model-oriented' nuclear physics approach towards real 
QCD descriptions. Recent theoretical investigations encourage such efforts 
by strongly pushing the limits of previous QCD-techniques towards
a much better description of semi-inclusive and exclusive reactions. 
This requires an extension of the classification of polarized twist-2 and
twist-3  distribution and fragmentation functions plus a realistic
phenomenology and more sophisticated hadron wave functions. A great potential 
to achieve an even deeper understanding of the nucleon structure may 
arise from a comprehensive, generalized analysis of many different
processes based on the new tool of skewed parton distributions
(SPDs). 

The study of hadron structure has another important facet in that it would 
supply badly needed input for the interpretation of data from 
Tevatron and LHC. A better understanding of the interplay between soft 
and hard contributions in exclusive processes is relevant for the success 
of the B-factories at hadron colliders. 
Issues of great importance for the LHC include a better 
determination of the gluon distribution for the whole range of Bjorken-$x$ as
well as a better understanding of isolated photon production, which is an 
important background for $H\rightarrow \gamma + \gamma$. 
 
Within the nuclear and particle physics communities there
exists an increasing conviction in the necessity of a new facility to study
polarized lepton-nucleon/nucleus scattering 
with very high luminosity and a high enough center-of-mass 
energy to cover a sufficient kinematic domain. 
This document outlines the TESLA-N project, which would use one arm of the
TESLA linear collider at DESY for a polarized electron-nucleon fixed-target 
experiment. The current 
discussions about ELFE@DESY, ELFE@CERN, eRHIC, EPIC, or a long-term 
high-energy option for CEBAF are all variations on the same subject.
TESLA-N is a highly competitive and very cost-effective alternative option. 
Its distinguishing property is the unique combination of large 
center-of-mass energy and high luminosity.

%==========================
\section{Physics Prospects}
%==========================

\enlargethispage{\baselineskip}
The HERMES results that have emerged over the recent past are 
demonstrating the richness of polarized electron-nucleon physics.
The higher energy and the much higher luminosity of TESLA-N are expected 
to again significantly enlarge the number and variety of observable effects 
as well as the precision with which they can be studied. Naturally,
today a theoretical understanding is only available for a part of this 
potential. Hence, the following list of topics illustrates rather than 
exhausts the physics potential of {TESLA-N}. 

Detailed projections for the statistical accuracy attainable in one of the 
TESLA-N key experiments, the precise measurement of the transversity
distributions, are given in the following section. Projections 
for all other topics are included in the sections following below, whenever 
available. All given projections are based on an integrated luminosity of 
100 fb$^{-1}$. This represents a conservative estimate for {\it one} year 
of data taking (cf. section~\ref{lumi}).

%-------------------------------------------------
\subsection{Transversity Distributions}
%-------------------------------------------------

The nucleon as a
spin 1/2 hadron is characterized by three independent flavor sets of
(leading order) quark distributions. The distributions 
$q(x,Q^2)$  - or $f_1^q(x,Q^2)$ - describe the 
unpolarized nucleon. The transversity distributions $\delta q(x,Q^2)$ 
- also referred to as $h_1^q(x,Q^2)$ or sometimes as $\Delta_T q(x,Q^2)$ 
- as well as the helicity distributions $\Delta q(x,Q^2)$ - also referred to 
as $g_1^q(x,Q^2)$ - describe aspects of the internal spin structure of the 
nucleon. One important difference between the latter two lies in their 
different QCD evolution. In contrast to the helicity distributions, 
the transversity distributions decouple from gluons.
The reason is, that the transversity distributions are chirally odd,
involving correlations  between left- and righthanded quarks.
Since $\Delta q(x,Q^2)$ and $\delta q(x,Q^2)$ describe the 
quark polarization in longitudinally and transversely polarized nucleons,
respectively, they are independent functions. However, in the 
most naive approximation using non-relativistic quarks 
$\delta q(x) \approx \Delta q(x)$ can be expected.

The first moments of the distribution functions give particular charges,
which are matrix elements of local operators. 
For the unpolarized distributions the
first moments of $q(x, Q^2)-\bar q(x, Q^2)$ give the flavor charges. 
For the helicity distributions the first moments of 
$\Delta q(x, Q^2) + \Delta \bar q(x, Q^2)$ give the axial charges 
$\Delta q(Q^2)$. The flavor sum of these axial charges, 
$\Delta \Sigma(Q^2)$, is the longitudinal quark spin fraction whose 
properties have given rise to all the commotion around the 
nucleon spin because of its anomalous evolution involving the polarized
gluon distribution. 
The first moments of $\delta q(x, Q^2) - \delta \bar q(x, Q^2)$ are
called tensor charges $\delta q(Q^2)$; their flavor sum is
denoted $\delta \Sigma(Q^2)$. Experimentally
nothing is known about the tensor charges, in contrast to the flavor
and axial charges. While for the axial charges the nonsinglet combinations
can also be measured in low-energy experiments (weak decays), no such
experiments are known for the tensor charges. 
The tensor charges $\delta q(Q^2)$ and their flavor 
sum $\delta \Sigma(Q^2)$ are valence objects and decouple from gluons
and sea quarks. In this respect, the tensor charges are expected to
be closer to the non-relativistic limit than the axial charges. 
This is supported by recent lattice QCD calculations \cite{aoki,capitani}. 
Reference~\cite{aoki} quotes values of $\Delta\Sigma= 0.18 \pm 0.10$ for
the longitudinal quark spin fraction and $\delta \Sigma= 0.562 \pm 0.088$ 
for the quark tensor charge at $Q^2 = 2$~GeV$^2$. 

As mentioned previously,
the transversity distributions $\delta q(x,Q^2)$ are not accessible in
inclusive measurements, because they are chirally odd and only occur
in combinations with other chirally odd objects.
In semi-inclusive DIS of unpolarized leptons off transversely
polarized nucleons several methods have been proposed to access 
$\delta q(x,Q^2)$ via specific single target-spin asymmetries:
\begin{enumerate}
\item An asymmetry that involves $\delta q(x,Q^2)$ in combination with the
  chirally odd polarized fragmentation function $H_1^{\perp(1)}(z)$ can be extracted 
from the azimuthal distribution of the produced single hadron 
\cite{collins,collins2,kotmul97,piet,boerm}.
This fragmentation function correlates the transverse spin of a quark with 
a preferred transverse direction for the production of the pion.
\item A measurement of the momenta of two leading pions gives access to an
asymmetry in which $\delta q(x,Q^2)$ combines with a so-called interference 
fragmentation function~\cite{intFF,intFF2,intFF3}. 
Here the transverse spin of the
quark is correlated with the relative transverse momentum between the pions.
\label{optii}
\item The determination of transverse components of the spin vector of 
produced $\Lambda$ particles allows the measurement of an asymmetry where 
$\delta q(x,Q^2)$ combines with a polarized fragmentation function 
$H_1(z)$~\cite{barone}.
\item Vector-meson production 
provides other ways to probe $\delta q(x,Q^2)$ employing polarimetry
and azimuthal asymmetries~\cite{bacmul,bamu}. For $\rho$-production, where
the polarimetry involves a pion pair, it is part of the above
two-pion production.
\end{enumerate}  
Option (1) offers the experimentally most direct access to 
$\delta q(x,Q^2)$. 
An appropriately weighted cross-section asymmetry can be expressed
as a flavor-sum where each transversity 
distribution function $\delta q(x,Q^2)$ enters in combination with a 
hitherto unknown polarized fragmentation function $H_1^{\perp(1)q}(z)$ 
of the same flavor~\cite{kotmul97}:
\begin{equation}
A_T(x,Q^2,z) =
 P_T \cdot D_{nn} \cdot
\frac{\sum_q e^2_q \  \delta q(x,Q^2) \ H_1^{\perp(1)q}(z)}
     {\sum_q e^2_q \  q(x,Q^2) \ D^q_1(z)}
\label{asy-QPM}
\end{equation}
Here $D_{nn}$ is the transverse polarization transfer
coefficient, $P_T$ is the nucleon's transverse polarization, and
$D^q_1(z) $ is the unpolarized quark fragmentation function that
recently has attracted renewed interest (cf. section~\ref{fragment}).

Measurements of different asymmetries in the production 
of positive and negative pions on proton and deuteron targets ($A_p^{\pi^+}$,
$A_p^{\pi^-}$, $A_d^{\pi^+}$, $A_d^{\pi^-}$)
allow the simultaneous reconstruction
of the shapes of the unknown functions $\delta q(x,Q^2)$ and the 
ratio $H_1^{\perp(1)}(z) / D_1(z)$. This ratio is considered to
be flavor independent in the context of this study.
The relative normalization can be fixed through independent measurements of
$H_1^{\perp(1)}(z)$, e.g. in $e^+e^-$ experiments. 
Alternatively, an additional assumption can be used, where one of the
possibilities is to relate $\delta q(x)$ to $\Delta q(x)$ at small values 
of $Q^2$. The differences between both are smallest in the region of
intermediate and large values of $x$, hence the 
normalization ambiguity can be resolved at $x_0=0.25$ by assuming~\cite{KNO}:
\begin{equation}
\label{h1eqg1}
\delta u (x_0, Q_0^2) = \Delta u(x_0, Q_0^2)
\end{equation}
Measurements of all possible asymmetries, $A_{p,d}^{\pi^+ (\pi^-)}$,
for $N_{(x, Q^2)}$ points in the $(x, Q^2)$-plane and for $N_z$ points in 
$z$ yield $4\cdot N_{(x, Q^2)}\cdot N_z$ measurements. This must be 
compared to $4\cdot N_{(x, Q^2)}$ unknown parameters, corresponding to
the quark distributions $\delta u(x, Q^2)$, $\delta d(x, Q^2)$, 
$\delta\bar{u}(x, Q^2)$, $\delta\bar{d}(x, Q^2)$, and to $N_z$ unknown values 
of $H_1^{\perp (1)}(z)/D_1(z)$. If kaon asymmetries are measured
in addition, the distributions $\delta s(x, Q^2)$ and 
$\delta \bar{s}(x, Q^2)$ can be included as well. 
This defines an overconstrained set of coupled equations which 
can be solved using a standard minimization procedure.

For the determination of the projected statistical accuracies for future
measurements of $\delta q(x,Q^2)$ at TESLA-N reasonable input is required
for the unknown functions $\delta q(x,Q^2)$ and 
$H_1^{\perp (1)}(z)/D_1(z)$. The former ones were assumed to coincide with
the GRSV LO parameterization~\cite{grsv96} for $\Delta q(x,Q^2)$ 
at the initial scale 
of $Q^2=0.4$ GeV$^2$  and evolved to higher values of $Q^2$ using the
DGLAP equations for transversity distributions. 
The resulting distributions approximately obey the Soffer 
bound~\cite{sbound,schaefer2}. All T-odd fragmentation functions
are constrained by a sum rule~\cite{schaefer3} but otherwise hardly
known. The ratio of fragmentation functions was modelled by adopting
the approach of reference~\cite{kotmul97} and fitting the
parameters to recent HERMES~\cite{hermes-azimuthal} and 
DELPHI~\cite{delphi-azimuthal} data. 
These results indicate that the fragmentation 
function $H_1^{\perp (1)}(z)$ may be quite sizeable.
Hadron distributions in semi-inclusive production were obtained using
the standard generators LEPTO~\cite{lepto65} and JETSET~\cite{jetset}.
\begin{figure}[htb]
\centering
\vspace*{-1.0cm}
\epsfig{file=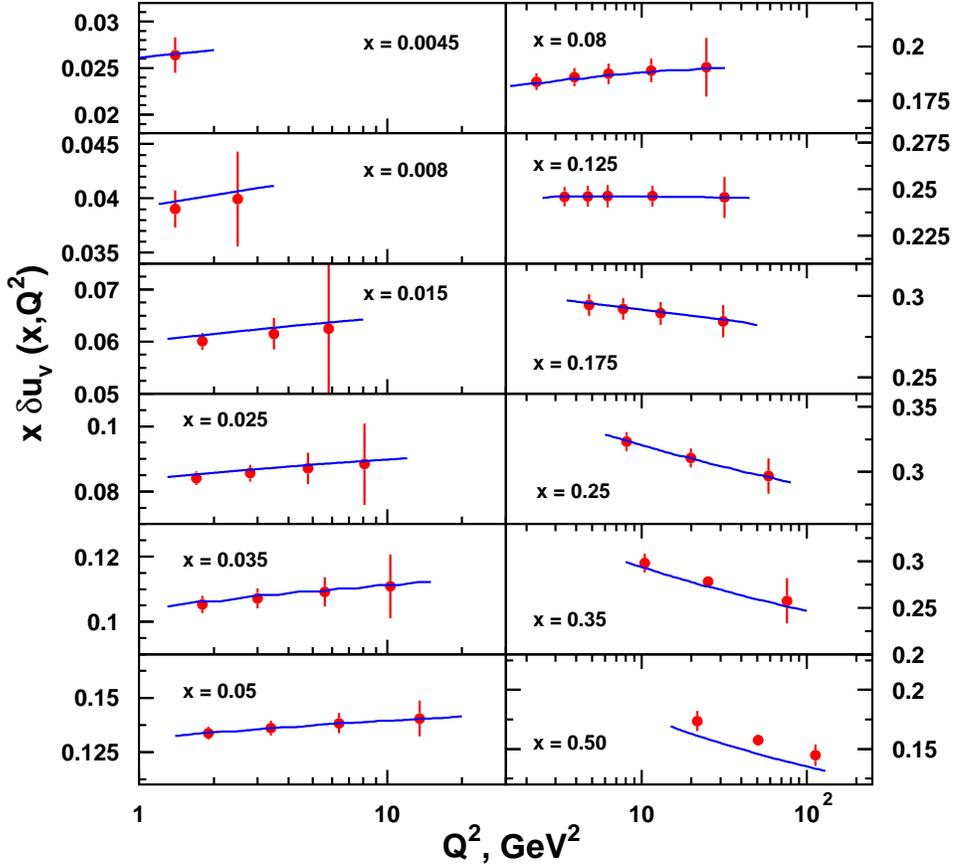,width=13.0cm}
\vspace*{-0.5cm}
\caption{\sf The valence $u$-quark transversity distribution as a function 
of $x$ and $Q^2$ as it would be measured at TESLA-N, based on an
integrated luminosity of 100 fb$^{-1}$. The curves show the LO
$Q^2$-evolution of the $u_v$-quark transversity distribution obtained 
with a fit to the simulated asymmetries.}
\label{uval-5mrad}
\end{figure}

The projected statistical accuracy for the measurement of the 
$(x, Q^2)$-dependence of the $u_v$-quark transversity distribution 
at TESLA-N is shown in figure~\ref{uval-5mrad}.
A broad range of $0.003 < x < 0.7$ can be accessed
in conjunction with $1 < Q^2 < 100$ GeV$^2$, with an impressive statistical
accuracy over almost the full range.
Because of u-quark dominance in pion electroproduction a somewhat lower
accuracy is attained in the reconstruction of the other transversity 
distributions, $\delta d_v$, $\delta \bar{u}$, and $\delta \bar{d}$.

\begin{figure}[htb]
\centering
\vspace*{-0.9cm}
\epsfig{file=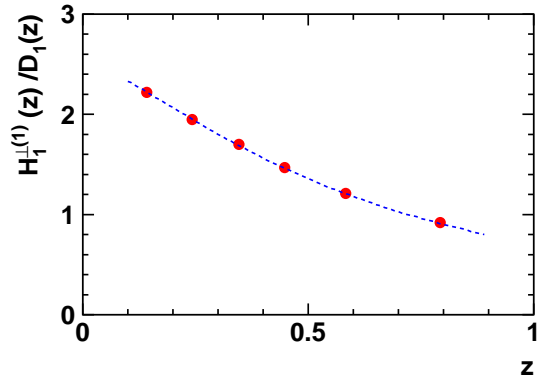,width=10.0cm}
\caption{\sf Projected accuracy of the ratio $H_1^{\perp(1)}(z)/D_1(z)$ 
of polarized and unpolarized fragmentation functions. The statistical
errors are smaller than the point size.}
\label{h1d1z-5mrad}
\end{figure}

There is a technically different approach 
to determine the unknown quark distributions
and fragmentation function ratios using parameterized transversity 
distributions. The starting point is a parameter-dependent ansatz for 
every $\delta q(x, Q_0^2 )$, e.g.
\begin{equation}
\label{anzats}
\delta q( x, Q_0^2) = \eta_q \cdot   
 x^{\alpha_q} (1 - x)^{\beta_q} (1 + \gamma_q x + \rho_q \sqrt{x})
\end{equation}
at a reference scale $Q_0^2$. Here $\eta_q$, $\alpha_q$, $\beta_q$, 
$\gamma_q$, and $\rho_q$ are free parameters. 
These free parameters and the unknown fragmentation function ratios
$H_1^{\perp (1)}(z)/D_1(z)$ are fitted to the simulated values of the 
asymmetries, calculated through eq.(\ref{asy-QPM}). In this procedure
the distribution functions are evolved in leading order to the necessary 
$Q^2$-values using the above ansatz. The resulting functional 
dependence for $x \cdot \delta u_v(x, Q^2 )$ is represented by the lines
drawn in figure~\ref{uval-5mrad}.
In addition, the fit also provides a projection for the accuracy 
of the tensor charges of $u$- and $d$-quarks of
$0.88 \pm 0.01$ and $-0.32 \pm 0.02$ at the scale of 1 GeV$^2$, respectively. 
Note that the absolute values of the tensor charges are defined  
to a large extent by the input distributions, although the values are rather
close to those predicted by lattice QCD calculations. 
At the same time, the fit yields precise values for the ratio of 
polarized and unpolarized favored quark fragmentation functions 
$H_1^{\perp(1)q}(z)/D^q_1(z)$. The projected accuracies, assuming 
flavor independence, are shown in figure~\ref{h1d1z-5mrad}. 

\begin{figure}[htb]
\centering
\vspace*{-1.0cm}
\epsfig{file=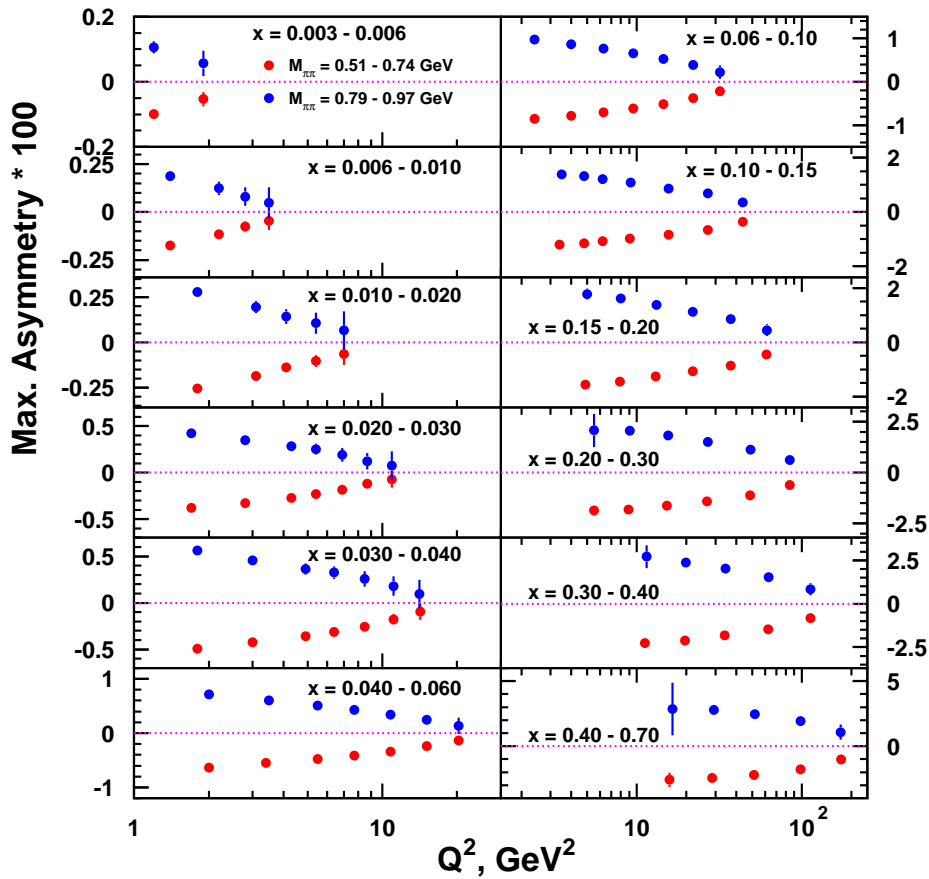,width=13.0cm}
\vspace*{-0.5cm}
\caption{\sf The maximum asymmetry for the two-pion system as a function
of $x$ and $Q^2$ as it would be measured at TESLA-N with an integrated
luminosity of 100 fb$^{-1}$. Results are shown separately for both two-pion 
mass regions.
}
\label{interf}
\vspace*{0.5cm}
\end{figure}

Option (2) (cf. p.~\pageref{optii}) focuses on the interference between the
$s$- and $p$-wave of the two-pion system around  the
$\rho$ mass.  Via the interference effect the
polarization information of the quark is contained in $\vec k_+
\times \vec k_- \cdot  \vec S_\perp$,  where
$\vec k_+$, $\vec k_-$, and $\vec S_\perp$ are the
three-momenta of $\pi^+$, $\pi^-$, and the nucleon's
transverse spin, respectively.
The corresponding asymmetry depends on the chirally odd 
$s-p$ wave interference quark fragmentation function
$\delta \hat{q}_{_I}(z)$ which is unknown at present, although it 
can be measured in $e^+ e^-$ experiments as well. 
Theoretically, there is an upper bound for this function that allows 
the estimation of the maximum possible asymmetry at TESLA-N.
The asymmetry is predicted to have different signs below and above the 
$\rho$-meson mass. To avoid averaging to zero, 
it must be considered separately in two regions 
of the two-pion mass, e.g. 0.51-0.74~GeV
and 0.78-0.97~GeV.
The corresponding expectations for the asymmetry are shown
in figure~\ref{interf}. 

\enlargethispage{\baselineskip}
At TESLA-N luminosity and kinematic range will be large enough to perform
a full flavor separation of both the distribution 
and the fragmentation functions of the transversely polarized nucleon.
This requires measurements of asymmetries in different final states, 
as well as $Q^2$-values that are large enough for factorization to be 
effective. TESLA-N will meet these requirements. 

%-------------------------------------------------
\subsection{Helicity Distributions}
\begin{figure}[htb]
\vspace*{-1cm}
\centering\includegraphics[angle=0, width=13cm]{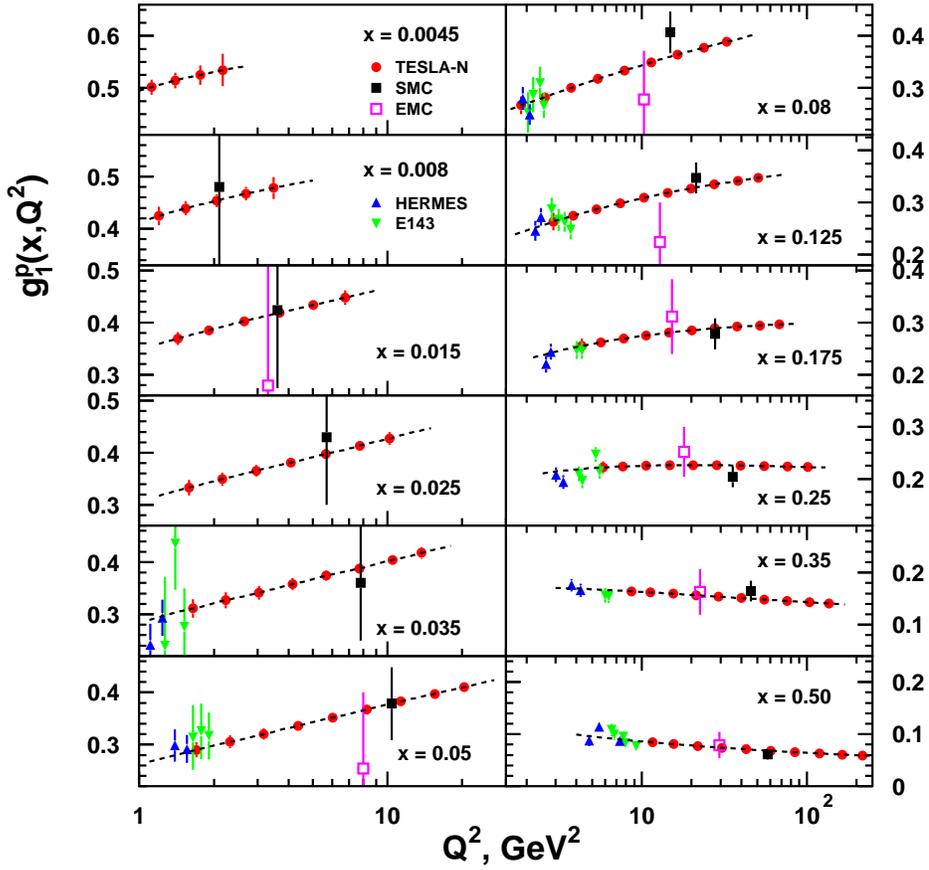}
\caption{\sf Projected statistical accuracy for a measurement of 
$g_1^p(x,Q^2)$ at TESLA-N, based on a luminosity of 100 fb$^{-1}$ and a 
minimum detector acceptance of 5 mrad. Two EMC/SMC data points are outside
of the shown vertical range.}
\label{figure:g1p5mrad}
\end{figure}

%--------------------------------------------------
The luminosity and kinematic range available at TESLA-N will allow
the determination of the longitudinal spin structure function $g_1(x,Q^2)$
through inclusive measurements with unprecedented accuracy. The
structure function $g_1(x,Q^2)$ represents a particular combination
of the helicity distributions $\Delta u$, $\Delta d$ and
$\Delta s$ and the corresponding antiquark distributions. The projection 
for $g_1^p(x,Q^2)$ is 
shown in figure~\ref{figure:g1p5mrad}. The anticipated precision 
in conjunction with the wide kinematic range will allow
studies that so far have not been possible. Prominent examples are
the determination of $\Delta G$ from NLO fits (cf. section~\ref{NLO}),
higher twist analyses (cf. section~\ref{twist}) and a precise determination
of the strong coupling constant $\alpha_s$ through the Bjorken sum rule.

SMC~\cite{smc-deltaq} and HERMES~\cite{hermes-deltaq} have provided 
relevant information on the longitudinally polarized u-quark and d-quark 
distribution functions. In addition, future HERMES data will allow 
to constrain $\Delta s(x,Q^2)$, $\Delta \bar u (x,Q^2)$ and
$\Delta \bar d(x,Q^2)$, however, with limited precision.
Semi-inclusive measurements with high precision can be provided at 
TESLA-N, due to the increased luminosity and kinematic range.

For example, a precise measurement of 
$\Delta \bar d(x,Q^2)-\Delta\bar u(x,Q^2)$ will strongly influence
the picture of the nucleon structure in general. This is the
direct parallel to the unpolarized case, where the difference
$\bar d(x,Q^2)-\bar u(x,Q^2)$ was measured to be large. An even
larger effect is actually predicted for 
$\Delta \bar d(x,Q^2)-\Delta\bar u(x,Q^2)$ by e.g. the 
chiral quark-soliton model~\cite{goeke}. 
The same holds true for the polarized strange-quark distributions
$\Delta s$ and $\Delta \bar{s}$ which has been an unresolved 
central issue in the discussion of the nucleon spin structure for 
more than 10 years.

%------------------------------------------------------------------
\subsection{Polarized Gluon Distribution}
%------------------------------------------------------------------
The polarized gluon distribution $\Delta G(x,Q^2)$ of the nucleon 
is essentially unknown as of today. There is a variety of approaches
to determine $\Delta G(x,Q^2)$; the most promising methods
in polarized DIS are the analysis of pairs of high-$p_{\perp}$ 
hadrons~\cite{bravhar}, and open charm production~\cite{reya}.

A first indication for the sign and approximate size 
of $\Delta G(x)$ has already been provided by HERMES through the analysis of
quasi-photoproduced pairs of `high'-$p_{\perp}$ hadrons~\cite{hermes-deltag}. 
However, the analysis has to rely on phenomenological event generators, 
which, due to the limited c.m. energy, are run at the limits of their 
validity range. The size of the resulting theoretical
error is controversial, but it is generally not expected that HERMES can 
provide a precision measurement of $\Delta G(x)$ along these lines. In 
contrast, for the considerably higher energies of TESLA-N these problems 
should be tractable. The anticipated COMPASS results will provide very 
valuable information, but a high precision experiment like TESLA-N is 
eventually needed for a reliable result. At RHIC, the determination of 
$\Delta G(x)$ is also not without problems in view of the great theoretical 
uncertainties in direct photon and heavy quark pair production and 
accounting for the fact that the detectors PHENIX and STAR are optimized 
for heavy-ion physics.  An independent determination in 
lepton-nucleon scattering is clearly needed to reach solid ground.
Up to now, no projections exist for measurements of the $Q^2$-dependence of  
the polarized gluon distribution. 
\begin{figure}[htb]
\vspace{-0.7cm}
\centering\includegraphics[angle=0, width=12cm]{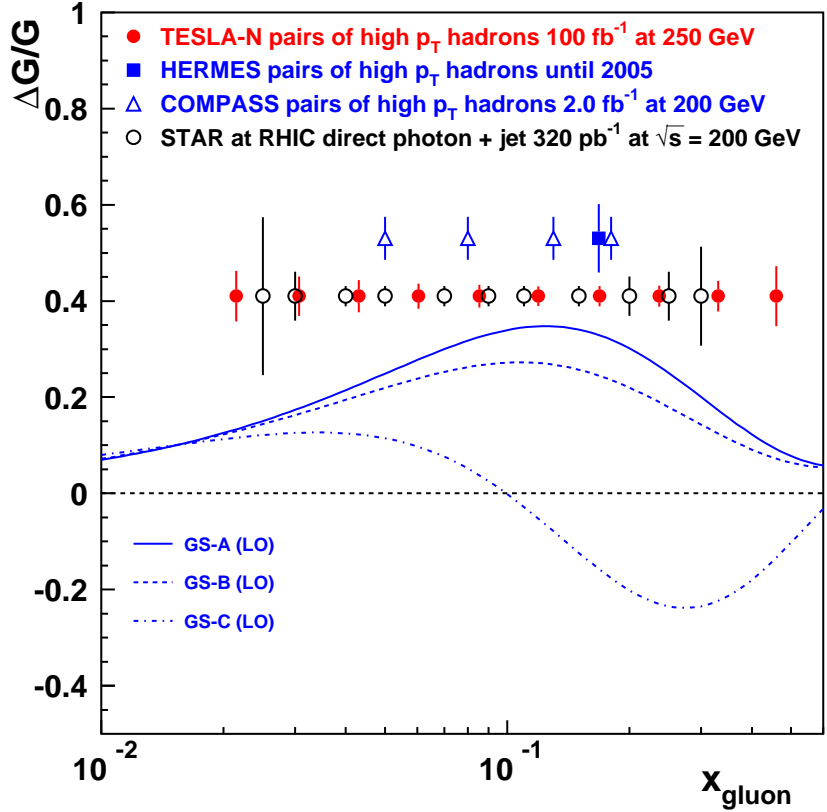}
\vspace{-0.5cm}
\caption{\sf Projected statistical accuracies for the measurement 
of $\Delta G(x)/G(x)$, based on an integrated luminosity of 100 fb$^{-1}$, 
in comparison to projections from RHIC~\cite{DG_G_rhic}, 
HERMES~\cite{DG_G_hermes}, and COMPASS~\cite{DG_G_compass}. 
A study of the systematic uncertainties due to the $x_{gluon}$ reconstruction 
procedure and due to QCDC background (right-most points) has not yet
been completed. The phenomenological predictions~\cite{GSLO} were calculated
for Q$^2$ = 10 GeV$^2$.}
\label{figure:deltag}
\end{figure}

The projected TESLA-N accuracy to measure $\Delta G(x)/ G(x)$ is shown in 
figure~\ref{figure:deltag}, in comparison to projected accuracies for HERMES, 
COMPASS and RHIC. 
In comparison to COMPASS, TESLA-N will have at least 50 times more statistics 
(cf. section~\ref{lumi}). Hence it will be the only envisaged polarized
lepton-nucleon scattering experiment capable to determine the ratio 
$\Delta G(x)/ G(x)$ over a wide range of $x$ with an impressive statistical
accuracy; systematic uncertainties have still to be studied (cf. caption). 
The overall $x$-range and the projected precision of the STAR measurement
at RHIC are comparable to the TESLA-N projection.
For completeness it has to be mentioned 
that high accuracy at large $x$ can also be realized at JLAB, but their 
`theoretical' systematic error can hardly be reduced below that at HERMES 
because of their low center-of-mass energy.  

In addition to the direct methods described above, QCD NLO fits
to the spin structure function  $g_1(x,Q^2)$ are able to yield
a parametric form of  $\Delta G(x,Q^2)$. However,  
no QCD fit to the existing data has yet been able to deliver a
statistically convincing determination of even the first moment 
$\Delta G(Q^2)=\int_0^1 dx \Delta G(x,Q^2)$.
At present, this {\it in}direct determination of $\Delta G(x,Q^2)$ 
is problematic, because at lower energies 
the effects of the evolution due to $\Delta G(x,Q^2)$ cannot be cleanly 
separated from higher-twist effects. 
A precision measurement of $g_1(x,Q^2)$ at {TESLA-N} will dramatically
enlarge the accuracy and the kinematic range,
as can be concluded from figure~\ref{figure:g1p5mrad}.
To obtain a projection for the first moment, 
a QCD NLO fit was performed in the $\overline{\mathrm{MS}}$ scheme using 
all DIS data published until summer 2000, giving a result of 
$0.43 \pm 0.21$ (stat.), at the 
scale of $Q^2=$1~GeV$^2$. The resulting structure function $g_1(x,Q^2)$, in its
parametric form, was then evolved into the kinematical region of 
\mbox{TESLA-N} 
and then used as additional input data for two new fits. Adding data that 
correspond to 100 fb$^{-1}$ using a proton target improves the statistical 
accuracy down to $\pm 0.06$. An additional data set obtained with
100 fb$^{-1}$ on a deuteron target yields a further improvement down to
$\pm 0.04$. This additional deuteron data set
considerably improves the statistical accuracy in the determination of the 
non-singlet quark distribution in the neutron, when comparing to existing 
data.\label{NLO} 

A comparison of this indirect determination of $\Delta G(x,Q^2)$ 
with the above described direct determinations will allow important 
consistency checks that in the end will lead to a reliable picture of how 
the gluons contribute to the nucleon spin. 

Last, but not least, results for the {\it un}polarized 
gluon distribution at large $x$ are of great importance to many 
searches for new physics and to the uncertainties in estimating conventional 
cross sections in the large-$x$ region for background processes to the 
Higgs-search at LHC. Present fits to the unpolarized gluon distribution in the 
region $x>0.15$ are still dominated by the old and partially inconsistent 
data of NA14/2, E691 and E687. The most suitable processes to determine 
the unpolarized gluon distribution at large $x$ are, as in the polarized
case, heavy-quark pair production and the production of pairs of 
high-$p_{\perp}$ hadrons. These measurements will automatically
also be available at TESLA-N.

%------------------------------------------------------------------
\subsection{Higher Twist}
%------------------------------------------------------------------ 

TESLA-N will be able to address a central issue of the present-day 
QCD discussions in inclusive and semi-inclusive physics, 
namely the role of higher twist.
It is clear that the applicability of perturbative QCD will eventually
come to an end for low photon virtualities due to the increase of
higher-twist effects. They hence play a crucial role in relating 
conventional perturbative QCD results to the bulk of hadron phenomenology. 
Today it is difficult to predict at which scales higher-twist 
contributions become important in the small-$x$ and large-$x$
domain of the different observables since the relevant parameters controlling
them are non-perturbative. Currently this is not even known for 
the well measured unpolarized 
structure function $F_2(x,Q^2)$. 
In addition, higher-order QCD corrections and 
higher-twist corrections cannot be dealt with 
independently~(cf. e.g.~\cite{claim,kataev1,kataev2,schaefer4}).
The knowledge of these corrections is also important for the presently 
available polarized data, which lie mostly in the $Q^2$-domain of only 
a few GeV$^2$. Obviously, a higher-twist analysis based on high precision
data for $g_1(x,Q^2)$ could help to clarify the situation substantially. 
This would also be important for 
spin physics in general, because it would reduce in present-day fits
the uncertainties due to neglected higher-twist contributions. \label{twist}

A precise measurement of the spin structure function $g_2(x,Q^2)$ remains a 
major challenge for future polarized DIS
experiments with transverse target polarization. The measurements
obtained so far~\cite{g2SLAC} will be improved 
by TESLA-N, extending the measurements down to $x$-values of 
5$\cdot$10$^{-3}$. 
Besides its twist-2 contribution $g_2(x,Q^2)$ contains a twist-3 term the 
isolation of which is important. 
At lowest order in QCD 
twist-2 and twist-3 contributions to $g_1(x,Q^2)$ and $g_2(x,Q^2)$ are 
connected by integral relations~\cite{WW,ELT,BT1,BT2} which can
be tested in this way. Moreover, if the $Q^2$-dependence of the
twist-3 contribution to $g_2(x,Q^2)$ can be isolated the validity of new QCD
evolution equations, cf. e.g.~\cite{BRAUN}, can be tested. Both issues 
provide new and important tests of QCD. 

There are several more distribution and fragmentation
functions for polarized electron-nucleon scattering.
Mulders and collaborators have given a 
classification of all twist-2 and twist-3 functions \cite{piet,boerm}. 
From a purist's point of view it can be argued that 
all of them are equally important, as they all test different features of 
nucleon structure and fragmentation dynamics. A more phenomenological point 
of view would be to concentrate on those that have an intuitive physics
significance or probe specific QCD dynamics. Presently a lot of  theoretical 
work is invested into the development of such an intuition 
(cf. e.g. \cite{mauro}).

%------------------------------------------------------------------
\subsection{Fragmentation Functions}
%------------------------------------------------------------------

\label{fragment}
A comprehensive study of fragmentation processes is of great value in
itself. To make full use of the data collected by the B-factories and 
(partly) LHC will require a good understanding of many different 
fragmentation processes. The high quality DIS data generated by TESLA-N 
would allow the fine-tuning of the fragmentation codes used for this purpose.
Contemporary semi-inclusive analyses usually assume
knowledge of the fragmentation functions, as obtained
from e$^+$e$^-\rightarrow$ hX, and use these as a tool  
in studying the parton distribution functions.
However, several new analyses of the e$^+$e$^-$ data have
appeared~\cite{leader2,leader3,leader4}. All agree very well with
the data, yet their derived fragmentation functions differ
significantly; in some regions of $z$ by 40-100 \%.
As a result, it has become crucial to use semi-inclusive DIS data
to measure parton distributions as well as fragmentation functions. There is
no problem of principle; all that is required is a sufficiently
large amount of high quality data~\cite{leader5}. While this is 
beyond present day experiments, TESLA-N should be able to stand
up to the challenge. However, no specific projections have been
worked out yet.

%------------------------------------------------------------------
\subsection{Specific Deuteron Structure Functions}
%------------------------------------------------------------------
\enlargethispage{\baselineskip}
In deep-inelastic scattering on a polarized spin~1 target 
new structure functions are involved that do not appear for a
spin~$\frac{1}{2}$ target. At leading twist the new functions are 
$b_{1(2)}(x,Q^2)$ \cite{manohar89} and $\Delta (x,Q^2)$ \cite{jaffe89}.
These hitherto completely unknown structure functions measure the extent 
to which the deuteron is not a trivial bound state of proton and neutron. 
$\Delta (x,Q^2)$ is especially interesting since it describes a flip of 
the photon helicity by two units. It probes the gluonic components of the 
deuteron wave function which cannot be identified with any contribution
from the constituent nucleons or virtual pions. 

The structure functions $b_{1(2)}(x)$ are accessible when the polarized 
electron beam is scattered off longitudinally polarized deuterons. The
measurement of $\Delta (x,Q^2)$ requires an unpolarized electron beam and
transversely polarized deuterons. In the latter case a characteristic 
azimuthal angular dependence of the cross section, $d\sigma \sim \cos{2\phi}$, 
is predicted. 

All these specific deuteron structure functions are expected to be of 
rather small size and thus a high luminosity polarized experiment as 
TESLA-N appears to be the ideal place to access information
on this non-trivial parton composition of the deuteron.

%
%===================================================================
\section{Layout of the Experiment}
%====================================================================
%
%---------------------------------------------------
\subsection{Polarized Target}
%--------------------------------------------------

One of the main ingredients of the TESLA-N apparatus is the polarized
target. To reach the required high luminosity with a small fraction
(20 nA, cf. section~\ref{beamsection}) of the total TESLA current 
(45 $\mu$A) a polarized solid 
state target of about 1 g/cm$^2$ areal density was chosen, similar in 
design to the one used at SLAC~\cite{E143-target}.

The polarized target will consist of a $^4$He evaporator cryostat, a 5~T 
Helmholtz-type magnet and a 140~GHz microwave system for permanent Dynamic 
Nuclear Polarization. The polarization is measured by Nuclear 
Magnetic Resonance. The maximum allowed heat load on the target is
limited by the cooling power of the evaporator cryostat to about 1~W at 
a temperature of 1~K. The total heat load on the target due to the
beam for a current of 20 nA has been calculated to be only about 
50 mW~\cite{steijger}. Hence, there should be no basic problem with the cooling.
Because 1~K is rather warm on the temperature scale 
of polarized targets, a strong magnetic field must be chosen to achieve 
reasonably high polarization values. The magnetic field is limited to 
5~T, because the power of microwave sources with frequencies higher than 
140~GHz is insufficient today. A symmetric Helmholtz design of the magnet 
combines excellent homogeneity with large opening angles for both, 
longitudinal and transverse polarization. The two main criteria for the 
choice of the target material are low dilution by unpolarized nucleons
and resistance against radiation damage with respect to the intense
TESLA beam. Therefore NH$_3$ ( $P_T$ = 0.8, f = 0.176)  
and $^6$LiD ($P_T$ = 0.3, f = 0.44) presently appear as the best choices 
to study electron scattering off polarized protons or deuterons.

A large number of physics questions can be addressed in high luminosity
running with different {\it unpolarized} nuclear targets. Targets with
very high atomic numbers can be easily constructed forming appropriately thin
foils. In this case electron beam currents may be possible 
that are considerably higher than 20 nA.

%-----------------------------------
\subsection{Polarized Electron Beam}
%-----------------------------------
\label{beamsection}
The electrons for TESLA-N will be accelerated together with the positrons 
in the north arm of the TESLA main accelerator. This `opposite charge option'
was chosen to be able to realize a separation between the beam for the 
eN-experiment and the main beam by a static magnet system. This system
would have a length of about 150 m and be located 
upstream of the separation for the two main interaction points 
(cf. figure~\ref{overview}). The beam energy initially will be 250~GeV;
energies up to 500 GeV may be possible in a later phase of TESLA.\\

\begin{figure}[h]
\centering
\fbox{\includegraphics[angle=0, width=12cm]{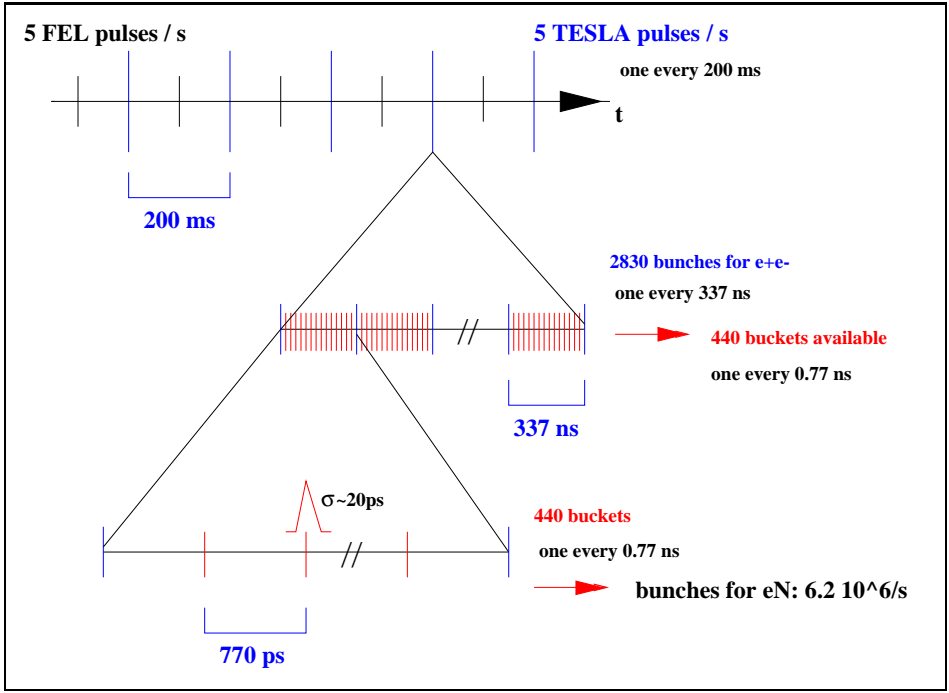}}
\caption{\sf Time structure of the polarized electron beam in TESLA
(south arm) fully exploiting the 1.3~GHz machine frequency.}
\label{figure:timestructure}
\end{figure}

Both TESLA arms are planned to run
with a 5 Hz macropulse repetition rate for e$^+$e$^-$-operation.
Additional 5 Hz will be run along a limited fraction of the south arm
to facilitate FEL operation.
The corresponding time structure of the polarized electron beam 
is illustrated in figure~\ref{figure:timestructure}.
The 0.5 \% north arm duty cycle in conjunction with the basic
machine frequency of 1.3 GHz has most severe consequences for the 
proposed experiment. Using the same time structure as foreseen for 
the e$^+$e$^-$-experiment, i.e. one bunch of 20 ps length crossing 
the target every 337 ns, would result in as much as a few hundred quasi-real 
photo-production events within these 20 ps.
This is an unacceptably high rate for an eN-experiment, because the scattered
electron must be uniquely assigned to the hadrons produced in the same
interaction. To minimize the number of multiple events per beam 
crossing while maximizing the luminosity, it is foreseen to fill every
bucket of the bunch train (one every 0.77 ns), while limiting the beam
current to 20 nA. This corresponds to 20k electrons per bunch and to
6.2 million bunches per second crossing the TESLA-N target. 

Although being beyond the scope of the present study, it should be
noted that a few improvements for eN-operation appear to be feasible.
\begin{enumerate}
\item Also along (part of) the north arm a 10 Hz macropulse repetition rate 
could be used. This would double all performance figures for e$^-$N-operation.
\item Two intermediate ejection points are technically feasible.
For the FEL, i.e. in the south arm, there will be 2 ejection points 
(at about 25 and 50 GeV). By RF tuning a dynamic range of about 2 will 
be routinely available, such that de facto energies 12.5 to 25 GeV (at 
point 1) and 25 to 50 GeV (at point 2) can be 'dialed'.
It is technically feasible to have two ejection points also in the north 
arm at e.g. 50 and 100 GeV. This would allow the selection of any energy 
between 25 and 100 GeV in addition to the full energy of 250 GeV.
\end{enumerate}
Physics requirements suggest to study e$^+$N-interactions as well.
It is technically unproblematic to install an additional (low intensity) 
positron source besides the separate electron source that is already
required for e$^-$N operation.
Since the `eN-positrons' will need a kicker magnet to be separated from the
'collider-positrons', only the extra 5 macro pulses in the 10 Hz 'a la FEL' 
mode could be used, thus limiting the duty cycle to 0.5\%.
Presently no solution is known to obtain polarized positrons in such a 
configuration. The production of polarized positrons requires $>$ 150 GeV 
electrons, as planned for TESLA e$^+$e$^-$ operation. However, at present 
it appears not realistic to assume that this system could also deliver 
polarized positrons for eN.

An electron current of 20 nA constitutes only about 0.04\% of the main beam 
current. Therefore the energy consumption for beam acceleration at TESLA-N
can be considered to be almost negligible. This advantage implies the drawback
that monitoring of the small electron beam cannot be done together with 
that of the high current beam in the main linac, but only before
and after acceleration. This requires further studies.

%-----------------------------------
\subsection{Overview of the Apparatus}
\begin{figure}[htb]
%\vspace{-0.5cm}
\includegraphics[angle=0, width=16cm, height=6cm]{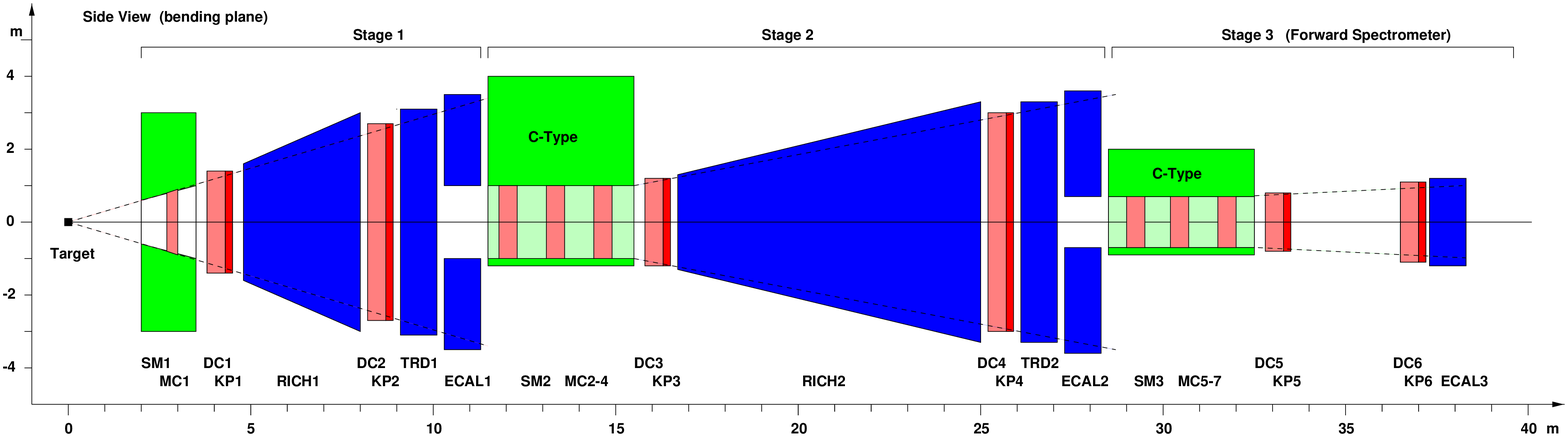}
\includegraphics[angle=0, width=16cm, height=6cm]{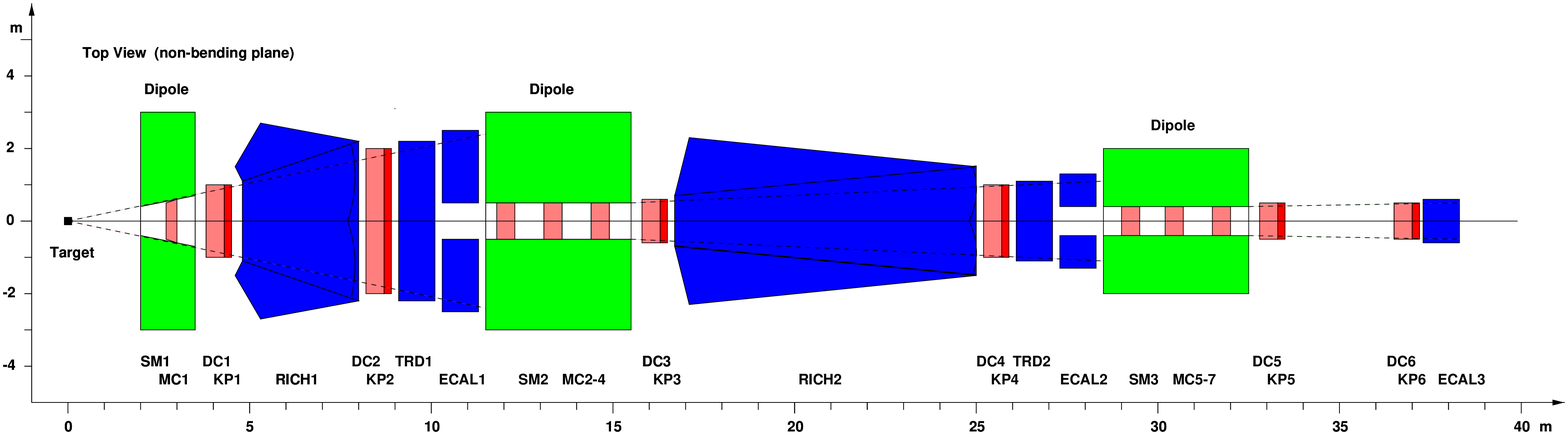}
\caption{\sf Schematic side view and top view of the envisaged TESLA-N 
apparatus. For an explanation of the acronyms see the text.}
\label{figure:sideview}
\end{figure}

%-----------------------------------
In a fixed-target electron-nucleon scattering experiment at 
250~GeV, acceptable resolutions in particle 
momentum and scattering angle may only be achieved by using a multi-stage 
spectrometer. A schematic sketch of a possible TESLA-N apparatus is
shown in figure~\ref{figure:sideview}. 
All three stages of the spectrometer will use large dipole 
magnets (SM1-3) for momentum analysis. As can be seen from the figure, the
overall dimensions of the TESLA-N apparatus are comparable to those of 
COMPASS \cite{compass-prop} because the kinematics of both experiments are 
similar.

Most hadrons are to be measured in Stage~1 , while
most of the scattered electrons and, in addition, a part of the 
leading hadrons will be detected in Stage~2. For both
Stage~1 and 2 electron/hadron separation, hadron identification, and
electron/photon separation will be very important and hence
their design looks similar to the HERMES spectrometer 
\cite{hermes-spectr}. Stage~3 is required to 
detect scattered electrons down to the lowest possible angles and will need 
adequate tracking capabilities combined with some electron/hadron
separation. 

While at COMPASS a thick target is traversed by incoming muons, the 
relatively thin solid state target planned for TESLA-N will
be hit by electrons that cause a much higher rate of bremsstrahlung.   
Its rate amounts to about 20\% of the incoming electron rate at a 
target thickness of 1 g/cm$^3$. Due to the magnetic deflection,
the resulting lower momentum electrons form a `sheet of flame' on their 
way down the spectrometer. While the width of the sheet-of-flame itself is
below 1~mm, its effective width corresponds to a possible wobbling area of the 
incoming electron beam that must be of the order of a few mm to match the 
target size. The electrons and the bremsstrahlung 
photons must not meet any material in their way to avoid background
showers. The safest way to ensure this is to provide a vacuum chamber 
that contains not only the high energy beam electrons, but also the
sheet-of-flame electrons and the radiated photons, as well as the 
synchrotron radiation produced in the three spectrometer magnets and
the target magnet. Instead of only a vacuum pipe an extended vacuum
vessel appears to be necessary. This vessel forms a narrow `slit'
whose height (in the bending plane) is increasing along the
spectrometer, while its width can be as low as $\pm$2 cm.  

The sheet-of-flame must be oriented towards a beam dump close to
the experiment. To this end, the dipole fields in SM1-3 should
be oriented horizontally and the above mentioned vacuum vessel
extends vertically down from the beamline. As a further
consequence, SM2 and SM3 will most likely be C-type magnets.
The integrated magnetic fields will be 2 Tm for SM1 and 5 Tm each
for SM2 and SM3. 

The envisaged very high luminosity of the experiment  
leads to very high interaction rates, so that a few hundred charged tracks
can be expected per `recorded event'. Here `recorded event' stands
for the pile-up of physics events over the typical integration
time of the tracking detectors (cf. section~\ref{lumi}). 
These conditions make it impossible for a single tracking 
device to have both the required very fast response and the necessary position
resolution. Therefore it is planned to combine fast tracking detectors (e.g.
scintillating fibres), so-called `key planes' (KP1-6), with precise 
tracking detectors. These detectors could be drift chambers  
(DC1-6). Similar to COMPASS, where a 2~ns resolution was shown 
recently \cite{DvHpriv1}, the fast but less precise detectors will serve to 
`snapshot' events on the bunch level which cannot be resolved by the slow 
but precise detectors (`fast-slow tracking'). 

The first section of SM1 will bend out the particles with
momenta below a few hundred MeV. Hence a first slow and less precise 
position detector (`magnet chamber',`MC1') may possibly be already 
accommodated within the gap of SM1. Since Stage~1 of the spectrometer 
will analyse predominantly particles in the momentum range of 1 to 
some 10~GeV, where multiple scattering is still an issue, the thickness of 
all Stage~1 tracking detectors must be optimized against their
precision and possibly a large low pressure container will be used.
Choosing 150 ns as a representative integration time for the `slow' detector,
the cell size of all detectors must be small enough to deal with the
expected high occupancies. The requirements for the Stage~2 detectors are 
expected to be less severe.

The electron-hadron separation in both Stage~1 and 2 will be provided by
combinations of transition radiation detectors with electromagnetic 
calorimeters (TRD1/2, ECAL1/2), complemented by ring-imaging Cerenkov 
detectors for hadron identification (RICH1/2). For Stage~3 only an
electromagnetic calorimeter (ECAL3) is foreseen. In addition, the gaps of 
SM2 and SM3 can be instrumented with suitable tracking detectors, 
e.g. proportional chambers (MC2-7), to minimize acceptance losses. 
Both ECAL1 and ECAL2 must not cover the entrance cone
to the next spectrometer stage, while all other detectors in Stage~1 and 2 
have to be designed with the central slit for the through-going flux of 
photons and high energy electrons, as discussed above.

Certain reaction channels greatly benefit from additional kinematic
constraints that can be obtained by measuring `recoil' particles. In the
given kinematics, recoil particles leave the target under laboratory polar
angles of a few tens of degrees. Their detection can hence be accomplished 
by a small-size barrel detector~\cite{steijger}
surrounding the target and/or forward 
`wheels' similar to those developed at HERMES \cite{hermes-wheels}. 
The target holding field may even allow for some momentum analysis, while 
some particle identification may be possible through ionization signals.

%--------------------------------------
\subsection{Luminosity and Acceptance}
%--------------------------------------
\enlargethispage{\baselineskip}
\label{lumi}
An areal target density of 1 g/cm$^2$ of polarizable material that is hit
by bunches carrying 20k electrons each, leads to a maximum possible
luminosity of 12~mb$^{-1}$ per bunch. With the above explained 6.2 million 
bunches per second this corresponds to a 
luminosity of $7.5 \cdot 10^{34}$ cm$^{-2}$ s$^{-1}$, which represents
the maximum value possible with the present TESLA design.

Table~\ref{table:lumi} shows the envisaged luminosity of TESLA-N in comparison
to other current or planned electron scattering facilities.
The first entry corresponds to the present TESLA design with a 5 Hz 
macropulse repetition rate in the north arm; the second entry applies
if a (technically feasible) rate of 10 Hz is assumed 
(cf. section~\ref{beamsection}).  
When comparing luminosities of fixed-target experiments (upper panel)
and collider experiments (lower panel), the degree of polarization and purity
(= fraction of the polarized material) of the involved nucleon have to
be taken into account. For example, when comparing a polarized
NH$_3$-target and a circulating proton beam,
the effective luminosity of the polarized fixed-target experiment is lower by
a factor of about 25.

\begin{table}[h]\centering
\begin{tabular}{|l|r|c|}
\hline
Experiment & c.m. Energy [GeV] & Luminosity [cm$^{-2}$ s$^{-1}$] \\
\hline
\hline
TESLA-N           &  22  \hspace*{1.5cm}  & $7.5\cdot 10^{34}$ \\
TESLA-N (10 Hz)   &  22  \hspace*{1.5cm}  & $1.5\cdot 10^{35}$ \\
COMPASS           &  20  \hspace*{1.5cm}  & $5.0\cdot 10^{32}$ \\
SLAC (incl.)      & 5 $\div$ 10 \hspace*{1cm}  & $5.0\cdot 10^{34}$ \\
HERMES (unpol.)   &  7.2 \hspace*{1.5cm}  & $4.0\cdot 10^{33}$ \\
HERMES (pol.)     &  7.2 \hspace*{1.5cm}  & $2.0\cdot 10^{31}$ \\
ELFE@CERN (unpol.) &  7  \hspace*{1.5cm}  & $1.0\cdot 10^{38} $ \\
ELFE@CERN (pol.)  &  7  \hspace*{1.5cm}  & $5.0\cdot10^{35} $ \\
\hline
HERA $\vec{e}\vec{p}$   & 318  \hspace*{1.5cm}  & $1.0\cdot 10^{31}$ \\
HERA eA           & 150  \hspace*{1.5cm}  & $1.0\cdot 10^{30}$ \\
eRHIC             & 100  \hspace*{1.5cm}  & $2.0\cdot 10^{32}$ \\
EPIC              &  31  \hspace*{1.5cm}  & $1.0\cdot 10^{33}$                    \\
\hline
\end{tabular}
\vspace*{0.3cm}
\caption{\sf Comparison of luminosities and c.m. energies
for current and planned electron scattering facilities}
\label{table:lumi}
\end{table}

Adopting a conservative ansatz for efficiencies, namely a combined up-time 
of accelerator and experiment of 0.33 in conjunction with an efficiency of 
the experiment of 0.75, results in the conservative overall efficiency
factor of 0.25. This factor leads to  maximum achievable 
integrated luminosities for TESLA-N of 1.6 fb$^{-1}$ per effective day, 
50 fb$^{-1}$ per effective months, and 600 fb$^{-1}$ per effective year. 
The term `effective' was chosen here to characterize a running period 
during which both accelerator and experiment operate routinely including 
all usually occurring day-by-day problems. 

At maximum luminosity every bunch (one every 0.77 ns) produces on average 
0.2 quasi-real-photoproduction events with $\nu>3$ GeV. 
For an average multiplicity of about 3 detected charged 
hadron tracks per physics event this corresponds to about 100 hadron tracks 
in Stage~1 per recorded event. The typical integration time and thus the 
length of the recorded event is assumed to be 150 ns, corresponding to about 
200 bunches. The additional rate from M\"oller electrons with an energy above 
1.5 GeV amounts to about 1 per bunch, or about 200 electron tracks per 
recorded event. However, M\"oller electrons reaching the 
spectrometer can be uniquely distinguished from DIS electrons with 
Q$^2 >$ 1 GeV$^2$ by their kinematics. 
Only about one DIS event with Q$^2 >$ 1 GeV$^2$, W$^2 >$ 4 GeV$^2$ 
and polar angles above 5 mrad will be contained in one recorded event. 

A crucial question for the analysis of DIS events is whether they can
be cleanly identified or whether they are mixed with other events.
For the above quoted 0.2 photoproduction events per bunch about 18\% of 
all DIS events will be accompanied by a photoproduction event produced 
by the same bunch. Off-line cuts on the total deposited energy, the track
multiplicity and the energy of the leading hadron have to be used to 
identify and remove these events. From preliminary considerations 
it is expected that in the end this multiple event fraction 
for DIS events can be safely reduced to a level of about 1\% or less. 

\enlargethispage{\baselineskip}
In certain areas more work has to be invested to solidify the assumptions
made above: \\
i) It is presently assumed that a time resolution of 0.77 ns can be realized
in the future for the fast tracking detectors at TESLA-N. As it was proven 
recently, today's technology already allows to reach 2 ns \cite{DvHpriv1}. 
In a conservative approach a beam current lower by a factor of 3 would have
to be assumed. \\
ii) The method to reduce the multiple event fraction in a DIS event from 18\% 
to the envisaged 1\% can only be developed on the basis of a careful Monte 
Carlo study. There is very little doubt that a factor of 3 can be realized
easily. In a conservative approach a beam current lower by a factor of 6 
would have to be assumed to arrive at the design value of 1\%. \\ 
It is anticipated that adequate answers can be found for these questions.
To leave a `safety margin' until the above questions will have been answered,
it was decided to assume for all physics projections a reduction of the beam 
current, i.e. consequently also of the luminosity, by a factor of 6. 
This decrease in beam current will relieve both point 
i) and ii). In the most conservative approach, where both i) and ii) 
are taken at their lower limits, the beam current and thus the luminosity 
for the physics projections needs not to be reduced further than the factor 
of 6, because another factor of 3 can be gained by running for three years 
instead of one. Altogether it thus appears to be a well-founded starting 
point that  100 fb$^{-1}$ per effective year is the conservative
integrated luminosity of TESLA-N.
This number was taken to calculate all projected 
statistical uncertainties throughout this document. It appears worth noting
that it is still a factor of 50 above the maximum achievable 
integrated luminosity of 2 fb$^{-1}$, presently planned for one
effective year of COMPASS running with the same overall efficiency
factor of 0.25. 

\begin{figure}[htb]\centering
\includegraphics[angle=0, width=15cm]{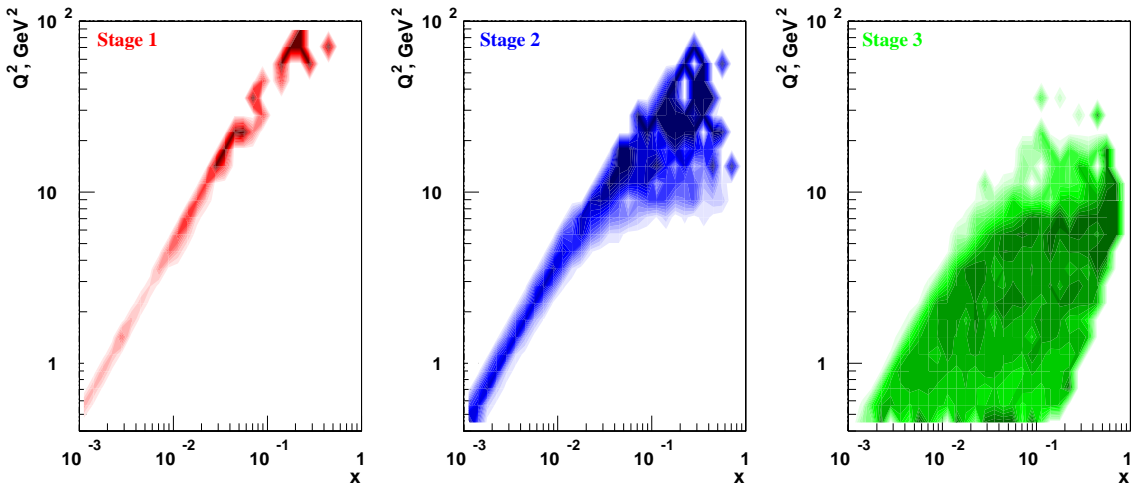}
\caption{\sf Acceptance of the TESLA-N spectrometer for the scattered electron.}
\label{q2acceptance}
\end{figure}

\begin{figure}[htb]\centering
\includegraphics[angle=0, width=12cm]{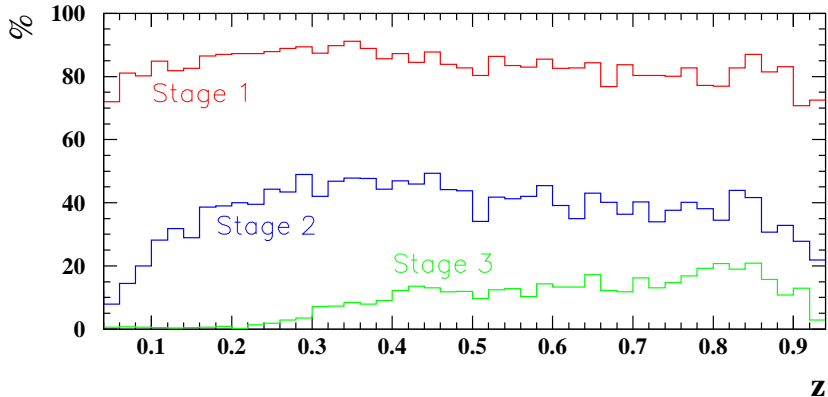}
\caption{\sf Acceptance of the TESLA-N spectrometer for the leading hadron.}
\label{zacceptance}
\vspace*{0.5cm}
\end{figure}

The acceptance of the TESLA-N spectrometer for the scattered electron
in the ($x,Q^2$) plane is shown in figure~\ref{q2acceptance}.
Electrons with high $Q^2$ ($>$10 GeV$^2$) are predominantly detected
in Stage~2, while low-$Q^2$ electrons ($<$10 GeV$^2$) are detected in
Stage~3. Figure~\ref{zacceptance} shows the acceptance for leading
hadrons as a function of $z=E_h/\nu$. More than 80\% of all leading
hadrons are detected in Stage~1 of the spectrometer while about
40\% of them are detected both in Stage~1 and 2 (for $z>0.15$). 
As a result, these hadrons are detected with good
momentum resolution independently of the vertex reconstruction.
The opening of SM1 limits the acceptance
to $\theta_x\leq 225$ mrad and $\theta_y\leq 280$.
The lowest possible detection angle $\theta_{min}$ is about 5 mrad
for momentum analysis within Stage~2 including the detection plane in front
of SM~2. If Stage~3 is used including the detection plane in front of
SM~3, $\theta_{min}$ can be reduced
to values as low as 2-3 mrad. 
These approximate figures are based on a width
of the vacuum vessel of $\pm$ 2 cm and a width of the DC frame next 
to it of 3 cm. For electrons,  $\theta_{min}$ directly
determines the lowest reachable Bjorken-$x$. 

%---------------------------------------------
\subsection{Resolution in Kinematic Variables}
%---------------------------------------------

Reliable values for the individual detector resolutions are not yet worked 
out. A momentum resolution on the level of 0.5 \% appears to be a
reasonable assumption. It can be achieved 
in spectrometer Stage~2 for tracks with momenta below 100 GeV, if a 
(realistic) position resolution of about 100 $\mu$m per space point is 
available. A similar momentum resolution for tracks crossing Stage~3 
with momenta up to 200 GeV requires better position resolutions. 
For the angular resolution 0.3 mrad can be assumed as a preliminary
value. The expected spread in the beam momentum (0.1\%)
is small enough to not affect the resolution in any of the kinematic 
variables. Possible beam energy losses prior
to the interaction have not been studied yet. 

The resolutions in the different kinematic variables are characterized by
two different effects. On the one hand, the resolution in $Q^2$ is dominated
by the resolution in the electron scattering angle. 
Only an angular resolution of the order of 0.3 mrad
or better can lead to $Q^2$-resolutions at the level of a few \% at large
$x$-values. None of the other spectrometer
resolutions have such a strong impact on the $Q^2$ resolution. 
On the other hand, the resolution in the variables $\nu, x$ and $z$ is
dominated by the momentum resolution of the spectrometer that, in turn,
has little impact onto the $Q^2$ resolution. 

Most of the non-leading and part of the leading hadrons will be detected 
in the Stage~1 of the spectrometer while the higher-momentum leading
hadrons will be measured both in Stage~1 and Stage~2.
A moderate {\em hadron} momentum resolution of the order of $1\%$ would 
be acceptable, provided that the electron momentum resolution is good enough.

%
%===================================
\subsection{Civil Engineering}
\begin{figure}[htb]
\centering
\epsfig{file=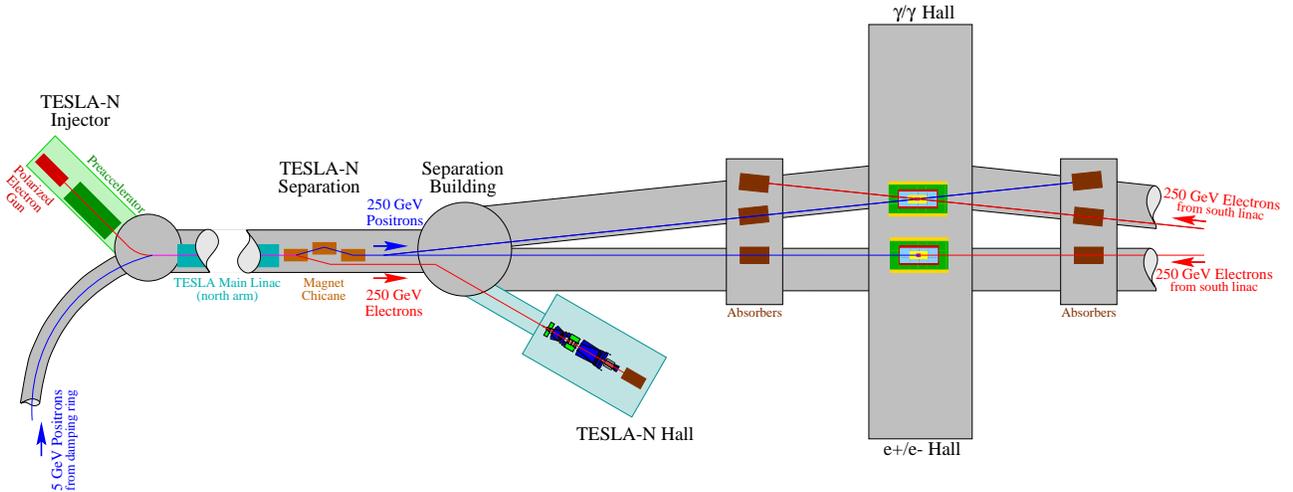,width=17.0cm}
\caption{\sf Schematic top view of the machine-related elements.}
\label{overview}
\vspace*{0.3cm}
\end{figure}

%===================================
%
The basic layout for the proposed
eN-experiment within the mostly fixed TESLA infrastructure is shown
in figure~\ref{overview}.

A separate electron gun system is required for TESLA-N at the north end of 
the TESLA machine.  It is envisaged to use a laser driven strained GaAs 
SLAC-type gun that must be made capable to deliver 20k highly polarized 
electrons per 0.77 ns. It must be followed by a separate preaccelerator whose
end energy and type are under discussion. Present options are a 
TESLA-type accelerator or a normal-conducting MAMI-type 
microtron. A short extra tunnel is required from the separation
building to the TESLA-N hall. The TESLA-N experimental hall would be
placed as far north as the site permits to minimize construction
costs. An extra beam absorber is required.

\vspace*{0.5cm}

%====================
\section{Summary}
%====================
%

This document presents the prospects for a polarized deep-inelastic
electron-nucleon scattering experiment at the TESLA facility at DESY.
For the first time a complete mapping of the $Q^2$- and $x$-dependence 
of both the helicity and the transversity distributions $\Delta q$ 
and $\delta q$ will become possible. 
Complemented by precise results on the polarized gluon distribution
most of the components of the angular 
momentum structure of the nucleon will be determined with high precision.
Hence, the measurements foreseen at TESLA-N will
constitute one of the most comprehensive and precise investigations 
of hadronic properties and tests of QCD techniques in the polarized sector.
These measurements will open an access to the hitherto unknown
chirally odd operators in QCD and thus greatly improve the
understanding of the role of chiral symmetry.

A possible layout for a fixed-target electron-nucleon scattering
experiment TESLA-N is presented as well. A separate hall is foreseen 
north of the main $e^+e^-$-interaction point. First design considerations 
for a polarized target, a three-stage spectrometer and a recoil detector are 
discussed. It is concluded that the experiment is technically feasible, 
although many aspects of the design require further study. 

The proposed deep-inelastic eN-experiment at TESLA constitutes a highly 
competitive and very cost-effective solution. It will be unique
as it combines high luminosity with large center-of-mass energies, 
using highly polarized electron beams and targets. 
The possibilities of using unpolarized targets and of experiments with a 
real photon beam turn TESLA-N into a versatile next-generation facility 
at the intersection of particle and nuclear physics.
Finally, this experiment would be the natural continuation of the HERA 
tradition at DESY in studying the structure of the nucleon with 
electromagnetic probes.

\newpage

%
%------------


\begin{thebibliography}{99}
\bibitem{schaefer1}
O.~Martin, A.~Sch\"afer, M.~Stratmann, W.~Vogelsang,
{\it Phys.Rev.} {\bf D60}, 117502 (1999)
\bibitem{hermes-LRP}
HERMES Collaboration, Hermes internal note 00-003, DESY, 2000. 
\bibitem{compass-prop}
COMPASS Collaboration, CERN/SPSLC 96-14 (1996). 
\bibitem{Jaffe:1997yz}
R.~L.~Jaffe, MIT-CTP-2685, hep-ph/9710465. 
\bibitem{Pobylitsa00} P.~V.~Pobylitsa et al.,
Bochum University preprint RUB-TPII-15/00 
\bibitem{aoki} 
S.~Aoki, M.~Doui, T.~Hatsuda and Y.~Kuramashi,
{\it Phys. Rev.} {\bf D56}, 433 (1997) 
\bibitem{capitani}
S.~Capitani et al., DESY~99-064, 1999 
\bibitem{collins}
J.C.~Collins, {\it Nucl. Phys.} {\bf B396}, 161 (1993)
\bibitem{collins2} 
J.C.~Collins, S.F.~Heppelmann, G.A.~Ladinsky,
{\it Nucl. Phys.} {\bf B420}, 565 (1994) 
\bibitem{kotmul97}
A.M.~Kotzinian, P.J.~Mulders, {\it Phys. Lett.} {\bf B406}, 373 (1997). 
\bibitem{piet}
P.J.~Mulders, R.D.~Tangermann, 
{\it Nucl. Phys.} {\bf B461}, 197 (1996) 
\bibitem{boerm}
D.~Boer and P.J.~Mulders, 
{\it Phys. Rev.} {\bf D57}, 5780 (1998) 
\bibitem{intFF} 
J.~C.~Collins and G.~A.~Ladinsky, hep-ph/9411444 
\bibitem{intFF2}
R.~L.~Jaffe, X.~Jin and J.~Tang, {\it Phys. Rev. Lett.}  
{\bf 80} (1998) 1166 
\bibitem{intFF3}
A.~Bianconi, S.~Boffi, R.~Jakob and M.~Radici,
{\it Phys. Rev.} {\bf D62} 034008 (2000)
\bibitem{barone} V.~Barone, 
{\it Nucl.Phys.} {\bf A666-667}, 282 (2000)
\bibitem{bacmul} X.~Ji, {\it Phys. Rev.} {\bf D 49}, 114 (1994) 
\bibitem{bamu}
A.~Bacchetta, P.J.~Mulders, VUTH 00-20, hep-ph/0007120
\bibitem{KNO}
V.A.~Korotkov, W.-D.~Nowak, K.A.~Oganessyan, 
DESY~99-176, hep-ph/0002268. 
\bibitem{grsv96}
M.~Gl\"uck et al., {\it Phys. Rev.} {\bf D53}, 4775 (1996) 
\bibitem{sbound}
J.~Soffer, {\it Phys. Rev. Lett.} {\bf 74}, 1292 (1995)
\bibitem{schaefer2}
O.~Martin, A.~Sch\"afer, M.~Stratmann, W.~Vogelsang,
{\it Phys.Rev.} {\bf D57}, 3084 (1998)
\bibitem{schaefer3}
A.~Sch\"afer and O.~Teryaev, {\it Phys.Rev.} {\bf D61}, 077903 (2000)
\bibitem{hermes-azimuthal}
HERMES Coll., A.~Airapetian et al., {\it Phys. Rev. Lett.} 
{\bf 84}, 9047 (2000)  
\bibitem{delphi-azimuthal}
A.V.~Efremov, O.G.~Smirnova, L.G.~Tkachev,
{\it Nucl. Phys. Proc. Suppl.} {\bf 74}, 49 (1999) 
\bibitem{lepto65}
G.~Ingelman, A.~Edin, J.~Rathsman, {\it CPC} {\bf 101}, 108 (1997) 
\bibitem{jetset}
T.~Sj\"ostrand, {\it CPC} {\bf 82}, 74 (1994) 
\bibitem{smc-deltaq} 
B.~Adeva et al., {\it Phys. Lett.} {\bf B420}, 180 (1998) 
\bibitem{hermes-deltaq}
HERMES Coll., K.~Ackerstaff et al., {\it Phys. Lett.} 
{\bf B464}, 123 (1999) 
\bibitem{goeke}
K.~Goeke, P.V.~Pobylitsa, M.V.~Polyakov, D.~Urbano, hep-ph/0003324 
\bibitem{bravhar} A.~Bravar, D.~v.Harrach, A.~Kotzinian, {\it Phys. Lett.} 
{\bf B421}, 349 (1998)  
\bibitem{reya} M.~Gl\"uck, E.~Reya, W.~Vogelsang, {\it Nucl. Phys.} 
{\bf B351}, 579 (1991) 
\bibitem{hermes-deltag}
HERMES Coll., A.~Airapetian et al., {\it Phys. Rev. Lett.} 
{\bf 84}, 2584 (2000)  
\bibitem{DG_G_rhic}
L.~Bland, hep-ex/9907058 
\bibitem{DG_G_hermes}
E.~Aschenauer, W.D.~Nowak, private communication 
\bibitem{DG_G_compass}
J.~Nassalski, {\it Acta Phys. Pol.} {\bf B29}, 1315 (1998) 
\bibitem{GSLO} 
T.~Gehrmann, W.J.~Stirling, {\it Phys. Rev.} {\bf D53}, 6100 (1996) 
\bibitem{claim}
U.K.~Yang and A.~Bodek, {\it Eur. Phys. J.} {\bf C13}, 241 (2000) 
\bibitem{kataev1}
A.L.~Kataev et al., {\it Phys. Lett.} {\bf B417}, 374 (1998)
\bibitem{kataev2}
A.L.~Kataev, G.~Parente, A.V.~Sidorov, hep-ph/0001096 
\bibitem{schaefer4}
E.~Stein, M.~Maul, L.~Mankievicz, A.~Sch\"afer,
{\it Nucl. Phys.} {\bf B536}, 318 (1998)
\bibitem{g2SLAC}
E155x Coll., P.~Bosted et al., 
{\it Nucl. Phys.} {\bf A663--664} (2000) 297. 
\bibitem{WW}
S.~Wandzura and F.~Wilczek, 
{\it Phys. Lett.} {\bf B72} (1977) 195. 
\bibitem{ELT}
A.V.~Efremov, E.~Leader, O.V.~Teryaev,
{\it Phys. Rev.} {\bf D55} (1997) 4307  
\bibitem{BT1}
J.~Bl\"umlein, A.~Tkabladze, {\it Nucl. Phys.} {\bf B553} (1999) 427 
\bibitem{BT2}
J.~Bl\"umlein, A.~Tkabladze,
{\it Nucl. Phys. {\bf B} Proc. Suppl.} {\bf 79} (1999) 541. 
\bibitem{BRAUN}
V.M.~Braun, G.P.~Korchemsky, and A.N.~Manashov, {\it Phys. Lett.}
{\bf B476} (2000) 455. 
\bibitem{mauro}
M.~Anselmino, F.~Murgia, DFTT-06-2000, hep-ph/0002120 
\bibitem{manohar89}
R.L.~Jaffe, A.~Manohar, 
{\it Nucl. Phys.} {\bf B321}, 343 (1989) 
\bibitem{jaffe89}
R.L.~Jaffe, A.~Manohar, 
{\it Phys. Lett.} {\bf B223}, 218 (1989) 
\bibitem{leader2}
S.~Kretzer, {\it Phys. Rev.} {\bf D62}, 054001 (2000) 
\bibitem{leader3}
B.A.~Kniehl, G.~Kramer, B.~P\"otter, hep-ph/0003297 
\bibitem{leader4}
L.~Bourhis et. al., hep-ph/0009101  
\bibitem{leader5}
E.~Christova, E.~Leader, hep-ph/0007303  
\bibitem{E143-target}
K.~Abe et al., {\it Phys. Rev.} {\bf D58}, 112003(1998) 
\bibitem{steijger} 
J.J.M.~Steijger, G.~v.d.~Steenhoven, NIKHEF 2000-013 
\bibitem{hermes-spectr}
K.~Ackerstaff et al., {\it NIM} {\bf A417}, 230 (1998) 
\bibitem{DvHpriv1}
D.~von~Harrach, private communication 
\bibitem{hermes-wheels}
HERMES Collaboration, Hermes internal note 97-032, DESY, 1997 




\end{thebibliography}
\end{document}